\documentclass[aps,prx,10pt,twocolumn,superscriptaddress,nofootinbib]{revtex4-2}

\usepackage{amsmath,amssymb,times,hyperref,graphicx}
\hypersetup{colorlinks=true, linkcolor=blue, citecolor=blue, filecolor=blue, urlcolor=blue}

\newcommand{\A}{\mathrm{A}}
\newcommand{\B}{\mathrm{B}}
\renewcommand{\c}{\mathrm{c}}
\renewcommand{\d}{\mathrm{d}}
\newcommand{\m}{\mathrm{m}}
\newcommand{\p}{\mathrm{p}}
\newcommand{\ch}{\mathrm{ch}}
\newcommand{\cm}{\mathrm{cm}}
\newcommand{\kt}{k_\mathrm{B}T}
\newcommand{\FH}{\mathrm{FH}}
\newcommand{\vecr}{\mathbf{r}}

\newcommand{\eg}{\textit{e.g.}, }
\newcommand{\ie}{\textit{i.e.}, }
\newcommand{\cf}{\textit{cf.} }

\begin{document} 

\title{Polymer-residue accessibility shapes sequence dependence of critical temperatures for phase separation} 

\author{J. Pedro de Souza}
\thanks{J.P.D. and B.S. contributed equally to this work. }
\affiliation{Omenn-Darling Bioengineering Institute, Princeton University, Princeton, New Jersey 08544, United States}
\affiliation{Department of Chemical and Biomolecular Engineering, University of California, Los Angeles, Los Angeles, California 90095, United States}

\author{Benjamin Sorkin}
\thanks{J.P.D. and B.S. contributed equally to this work. }
\affiliation{Princeton Center for Theoretical Science, Princeton University, Princeton, NJ 08544, USA}

\author{Amala Akkiraju}
\affiliation{Department of Chemical and Biological Engineering, Princeton University, Princeton, New Jersey 08544, United States}

\author{Athanassios Z. Panagiotopoulos}
\affiliation{Department of Chemical and Biological Engineering, Princeton University, Princeton, New Jersey 08544, United States}

\author{Howard A. Stone}
\email{hastone@princeton.edu}
\affiliation{Department of Mechanical and Aerospace Engineering, Princeton University, Princeton, New Jersey 08544, United States}

\begin{abstract}
Biological polymers, such as intrinsically disordered proteins, play a central role in cellular biology, including mediating phase separation and controlling activity of biological condensates. The physical properties and functions of biopolymers are determined by their residue sequence. Recently, significant computational and theoretical efforts have been devoted to characterizing the combinatorially complex sequence dependence of biopolymer phase diagrams. Here, we quantitatively show that monomer accessibility is central to determining the strength of pair interactions. We formulate an analytical perturbative approach, phenomenologically precluding two polymers' centers of mass from overlapping within a correlation hole. This theory yields the correction to the strength of mean-field interactions  in terms of a residue-accessibility parameter (RAP), which accounts for the limited availability of inner monomers to interactions. RAP rationalizes the variations in critical temperatures found in extensive Monte Carlo simulations for thousands of two-letter polymer solutions of varying lengths and sequences. RAP may thus be effective for deciphering the polymer-sequence dependence of phase diagrams given any polymer length, set of monomer types, and polymer mixtures.
\end{abstract}

\maketitle

\section{Introduction}\label{sec:introduction}

Across living systems, the molecular grammar of biopolymers maps to their function via the sequence of their building blocks~\cite{ZarinElife21,RuffMolCell2022,hoffmann2025synapsin,WangCell2018}. Precise sequence features control conformational structure and the chemical binding of substrates~\cite{WangCell2018,hoffmann2025synapsin,PatilCell2023}. Furthermore, a wide variety of biopolymers, most prominently intrinsically disordered proteins, undergo phase separation to form biomolecular condensates, which organize biochemical reactions in membraneless organelles~\cite{MensahNat2023,BananiNRCB17,ShinSCI17}. In these often-disordered assemblies of biopolymers, how does sequence matter?

This question has been tackled  through a combination of theoretical formulations and simulations in which all inputs are controlled. Conventional mean-field models, such as the Flory--Huggins (FH) free energy~\cite{DoiBOOKpolymer}, average over all chain conformations, consequently predicting that only the overall polymer composition, not sequence, determines their phase behavior. However, molecular simulations resolving monomer identities reveal strong sequence dependence~\cite{panagiotopoulos2024sequence}. Moreover, even chains with identical monomer composition exhibit different critical temperature and densities for the phase transition across different sequences~\cite{panagiotopoulos2024sequence,RanaJCP2021,LiPRE2025, Akkiraju2025}. These deviations remain challenging to predict due to phase-space dimensionality scaling exponentially with polymer length. 

Experiments have probed the design rules driving phase separation through sequences~\cite{QuirozNatMat2015,martin2018relationship,hoffmann2025synapsin}. For example, a scrambling of sequence was found to modify the phase behavior of the intrinsically disordered region of synapsin-1~\cite{hoffmann2025synapsin}.  Other studies have  modified the sequence of proteins and their fragments by point mutations in order to identify which residues drive phase separation, with less attention to maintaining overall monomer compositions. A general outcome is that the sequence dependence has been linked to the precise chemical residues driving phase separation through intermolecular (electrostatic, cation-$\pi$, and $\pi$-$\pi$) interactions~\cite{martin2018relationship, bremer2022deciphering}. 

To explain the breadth of experiments and molecular simulations, a variety of theoretical approaches have been proposed. These include single-chain predictors, such as the sequence charge~\cite{sawle2015theoretical} and hydropathy~\cite{zheng2020hydropathy} decoration parameters, which  track with the ``blockiness'' of individual chains, in turn affecting their conformation. Although these predictors correlate with critical phase-separation parameters~\cite{RanaJCP2021,LiPRE2025}, as single-chain properties, they do not capture the multi-polymer physics underlying phase separation, with the correlation instead arising because both depend on, \eg blockiness but through distinct microscopic mechanisms. A more formal, sequence-feature description identifies individual sequence motifs and labels them as having unique interactions~\cite{patyukova2021phase}. More sophisticated, statistical-mechanical approaches, such as the random-phase approximation (RPA)~\cite{lin2017charge,lin2017random} and self-consistent field theory (SCFT)~\cite{PattersonMACRO2019,patterson2020monomer}, imbue  sequence dependence, albeit not in a readily applicable closed form and often without including inter-chain correlations.  %Advanced self-consistent field theory~\cite{patterson2020monomer} have also been used to understand the role of sequence on microphase separation.

Instead, we aim in the present work to highlight and theoretically quantify the role of \emph{residue accessibility} in phase separation. 
If a residue is ``buried'' near the center of a disordered chain, it will be less accessible for interactions with other polymers. The inverse also applies to residues at chain ends. These features are neither directly captured by single-chain measures nor by feature descriptions, and they are difficult to extract from more complex field theories. Nevertheless, previous studies have rationalized phase behavior through similar qualitative arguments based on the solvent exposure of particular residues~\cite{poudyal2023intermolecular}.

Here we make a previously unrecognized connection that the correlation hole~\cite{BOOK:deGennes,WangMACRO2017} affects the pair-interactions between polymers through variable accessibility of individual monomers for interactions. To make that proposal concrete, we obtain the first correction to the critical temperature due to the correlation hole within the widely used FH model. The key output of the model is a residue-accessibility parameter (Eq.~\eqref{eq:RAP}) that, when compared to Monte Carlo simulations where all inputs and sequence characteristics are controlled, strongly correlates with the critical temperature of thousands of two-letter chains. We expect this approach to be useful for predicting the role of accessibility in the interactions of intrinsically disordered proteins and other sequence-resolved polymers.

%Here, we include residue accessibility within a perturbed mean-field framework. We demonstrate that sequence-dependent phase behavior naturally arises owing to the correlation hole~\cite{BOOK:deGennes,WangMACRO2017} between neighboring interacting polymers. The key output of the model is a residue-accessibility parameter (Eq.~\eqref{eq:RAP}) that, when compared to Monte Carlo simulations where all inputs and sequence characteristics are controlled, strongly correlates with the critical temperature of thousands of two-letter chains. We expect this approach to be useful for predicting the role of accessibility in the interactions of intrinsically disordered proteins and other sequence-resolved polymers.

\section{Theory} \label{sec:theory}

We begin with the FH free energy~\cite{FloryJCP42,HugginsJCP41}, $F$, which is a conventional mean-field description for phase-separating polymer solutions. It combines the ideal entropy of mixing and an effective pair-interaction energy~\cite{DoiBOOKpolymer},
\begin{equation}
    \frac{v_\m F(T,\phi)}{V\kt}=\frac{\phi}{N}\ln\phi+(1-\phi)\ln(1-\phi)-\chi(T)\phi^2,\label{eq:FH}
\end{equation}
where $\phi$ is the total polymer volume fraction, $T$ is temperature, $k_\mathrm{B}$ is Boltzmann's constant, $V$ is the system volume, and $v_\m$ is a monomer's molecular volume. While the entropic contribution simply depends on the number $N$ of monomers in a chain, the interaction magnitude $\chi(T)\sim1/T$ (setting the energetic contribution) depends on polymer composition, although not on sequence. For instance, in lattice models with an inert effective solvent~\cite{DoiBOOKpolymer},
\begin{equation}
    \chi(T)=\frac{z\bar\epsilon}{2\kt},\qquad\bar\epsilon\equiv\frac1{N^2}\sum_{i,j=1}^N\epsilon_{t_it_j},\label{eq:chi_id}
\end{equation}
where $z$ is the coordination number, $\epsilon_{tt'}$ denotes the interaction strength between monomers of types $t$ and $t'$, and $t_i$ is the monomer type (``letter'') in the $i$th chain position. Within this framework, the critical volume fraction and interaction strength, respectively, 
\begin{equation}
    \phi_\c=\frac1{1+\sqrt{N}}\quad \hbox{and}\quad \chi(T_\c)=\frac12\left(1+\frac1{\sqrt{N}}\right)^2, \label{eq:chi_id_Tc}
\end{equation}
are sequence independent. Combining Eqs.~\eqref{eq:chi_id} and~\eqref{eq:chi_id_Tc}, the FH theory predicts a mean-field critical temperature $T_{\c,\FH}$,
\begin{equation}
    T_{\c,\FH}=\frac{z\bar{\epsilon}}{k_B}\left(1+\frac1{\sqrt{N}}\right)^{-2}.
\end{equation}
Here, $T_{\c,\FH}$ displays any composition dependence through $\bar\epsilon$, which motivated us to introduce sequence dependence into the energy term by means of a modified $\chi(T)$. 

Within the FH model, a modification of $\chi(T)$ will not yield a change in critical volume fraction. (Moreover, the FH prediction $\phi_\c\sim N^{-1/2}$ is often inadequate; \eg lattice simulations of long-homopolymer solutions display $\phi_\c\sim N^{-0.4}$~\cite{PanagiotopoulosMACRO1998}. Therefore, modifying the entropic term or considering the critical volume fraction within our current framework would be ineffective.) Based on Eq.~\eqref{eq:chi_id}, we define the rescaled critical temperature,  
\begin{equation}
    \tilde{T}_\c=\frac{T_\c}{T_{\c,\FH}}=\frac{\kt_\c}{z\bar{\epsilon}}\left(1+\frac{1}{\sqrt{N}}\right)^2,\label{eq:Tctilde}
\end{equation}
which regularizes for all composition dependence in the mean-field limit, \ie $\tilde{T}_\c=1$ if FH theory perfectly describes the system. Otherwise, $T_\c$ would deviate from the FH prediction in a sequence-dependent manner.

\begin{figure*}%[tbhp]
\centering
\includegraphics[width=.82\linewidth]{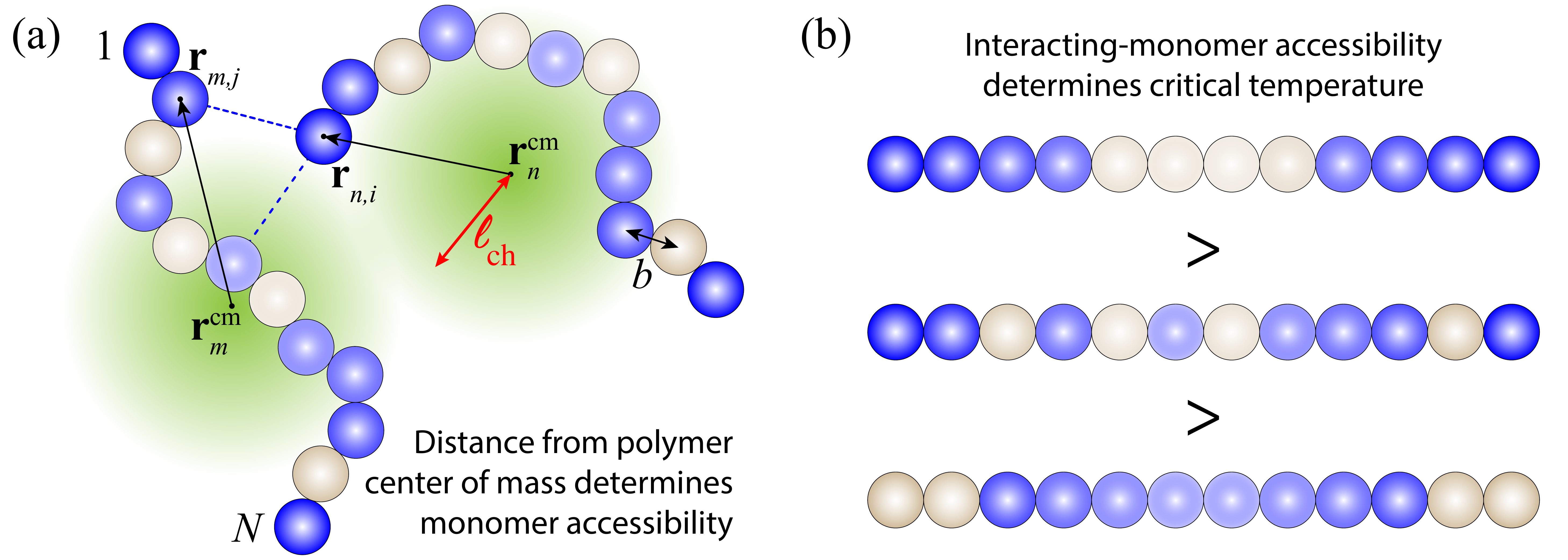}
\includegraphics[width=.82\linewidth]{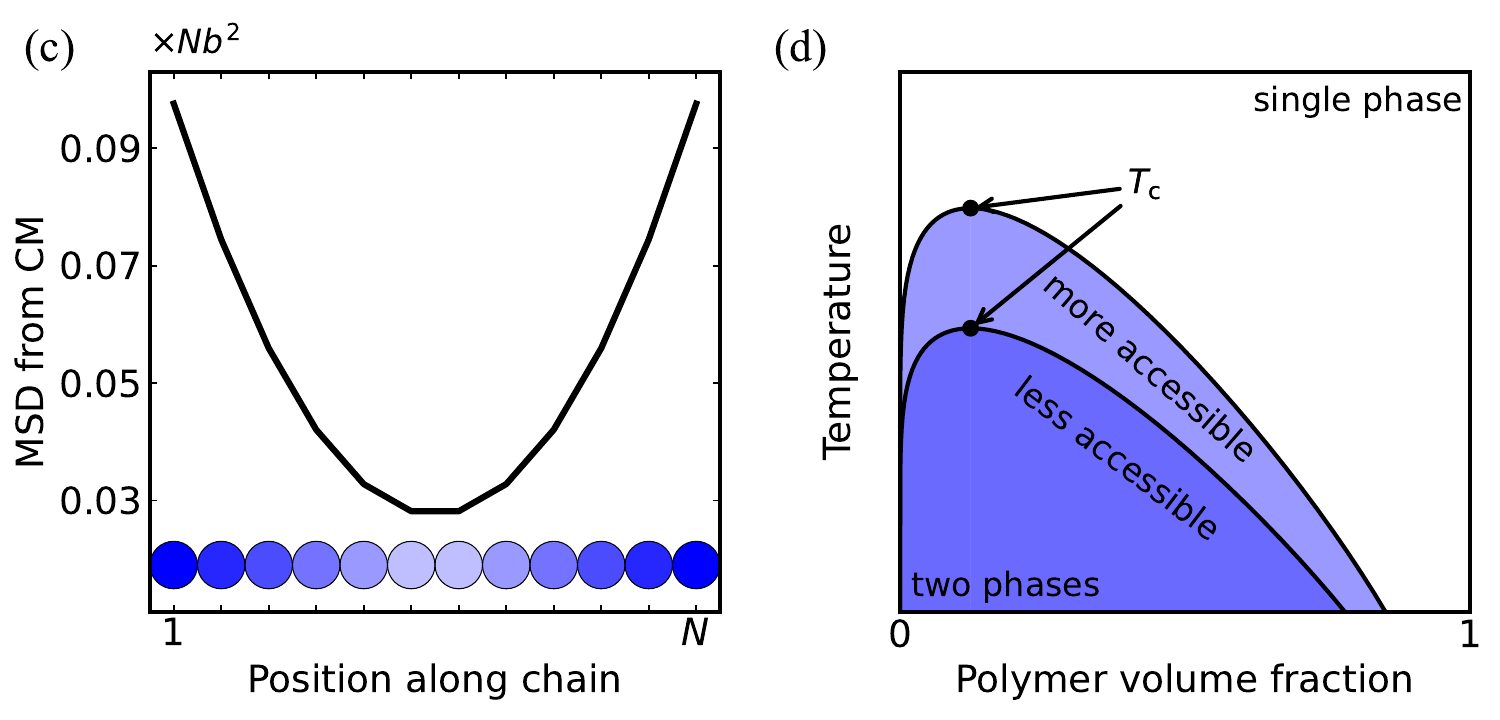}
\caption{Modeling the effect of residue accessibility on the critical temperature. (a) Overlap between the centers of mass (CMs) of two polymers is penalized when they approach within each other’s correlation hole. Each polymer consists of residues indexed $i=1,\ldots,N$ in a specified sequence (\eg blue beads represent attractive monomers, and beige represents inert monomers). The Kuhn length is denoted $b$. The spatial position of the $i$th monomer of the $n$th polymer is denoted $\vecr_{n,i}$. The $n$th polymer's CM, $\vecr_n^\cm$, is determined by the mean of the $N$ monomer positions. Since polymer CMs are limited to beyond the correlation hole, inner monomers are statistically less accessible for interactions than peripheral ones, as indicated by the bead color's shade. (b) When more attractive monomers are placed at the polymer periphery, they are more accessible for interactions with other polymers under the correlation-hole penalty. Stronger pair-interactions increase the critical temperature for phase separation. Illustrated are three polymer sequences with ranked critical temperatures. (c) Mean-squared distance (MSD) of a monomer from its host polymer's CM for a Gaussian chain versus monomer position, Eq.~\eqref{eq:MSD}, determining monomer's accessibility. (d) Schematic phase diagram for phase separation. The shaded regions correspond to the phase-coexistence regime for each of the two polymers solutions; the complementary regions indicate a single-phase regime. The region bound by a higher critical temperature corresponds to a polymer with more accessible attractive monomers.}
\label{fig:illust}
\end{figure*}

To impart sequence dependence to $\chi(T)$, we turn to the microscopic expression for the total pair-interaction energy among $N_\p$ polymers in the system, $(1/2)\sum_{n\neq m}^{N_\p}\sum_{i,j=1}^N\langle U_{t_it_j}(\vecr_{n,i}-\vecr_{m,j})\rangle$, where $U_{tt'}(\vecr)$ is the microscopic pair potential between $t$- and $t'$-type monomers separated by $\vecr$, $\vecr_{n,i}$ is the lab-frame spatial position of the $i$th monomer of the $n$th polymer, and $\langle\rangle$ denotes an ensemble average with respect to the Boltzmann weight; see Fig.~\ref{fig:illust}(a) for notation. The pair potential $U_{tt'}(\vecr)$ occurs within a length scale $\ell_U$. In mean-field theory, the polymers are taken to be statistically independent. Thus, in coexisting bulk phases, the polymer density varies slowly (\ie beyond the range of the pair potential, $\ell_U$), so one obtains Eq.~\eqref{eq:chi_id} with
\begin{equation}
    z\epsilon_{tt'}=-\frac1{v_\m}\int\d\vecr U_{tt'}(\vecr);\label{eq:epsilon}
\end{equation}
see details in Appendix~\ref{sec:deriv1}. Indeed, since the polymers are assumed independent, monomer interactions are insensitive to their chain position and the resulting $\chi(T)$ is sequence independent. To capture sequence dependence, polymer--polymer correlations must be included.

One manifestation of polymer correlations is the correlation hole~\cite{BOOK:deGennes,WangMACRO2017}. It is an emergent repulsion among the centers of mass (CMs) of two chains, arising from accumulated excluded-volume interactions between their constituent monomers. As a result, polymers are not independent but behave as weakly repulsive ``clouds'' (Fig.~\ref{fig:illust}(a)). This effect can be measured using scattering experiments for polymers with isotope-labeled monomers~\cite{BOOK:deGennes,BOOK:Higgins}, and is found ubiquitously in many models and polymer types~\cite{YatsenkoPRL04,ClarkJCP13,WangMACRO2017} (including that we consider in the following; \cf Fig.~\ref{fig:rdf_and_tp}(a)).

With depletion among polymer clouds in mind, we postulate that the strength of a pair-interaction between monomers on different chains depends on their position along each chain; namely, since the polymer CMs must be in closer proximity for the innermost monomers to interact, these monomers are statistically less accessible for interactions. A monomer's accessibility is related to its distance from the host chain's CM (Fig.~\ref{fig:illust}(c)). As such, if attractive monomers are less accessible, then the interactions that would have maintained a dense phase are weaker and the critical temperature would decrease compared to more interaction-accessible polymers (Fig.~\ref{fig:illust}(b,d)).

To make our reasoning quantitative, we define a polymer repulsion kernel $K(\vecr)$, describing how strongly the correlation hole is repelling two polymer CMs at separation $\vecr$. We define it as a unitless function (representing an occupied volume fraction), decaying from some value $K(0)\leq1$ to $K(|\vecr|\gg\ell_\ch)=0$ beyond the scale of the correlation hole, denoted $\ell_\ch$. We introduce $K(\vecr)$ into the conventional mean-field calculation perturbatively~\cite{BOOK:Hansen}. That is, once the separation between the two polymers' CMs is specified, we approximate the internal structure of both polymers to remain uncorrelated. Thus, each polymer follows the isolated-chain statistics, wherein the $i$th monomer's distance $\vecr'$ from its host chain's CM is drawn independently from a particular, prespecified, single-chain distribution $p_i(\vecr')$. Then, a configuration with a particular separation between the two polymer CMs is penalized by a factor $1-K(\vecr')$. With these approximations, Eqs.~\eqref{eq:chi_id} and~\eqref{eq:epsilon} are modified  to
\begin{multline}
    \chi(T)=-\frac1{2v_\m\kt}\frac1{N^2}\sum_{i,j=1}^N\int\d\vecr\int\d\vecr_1\int\d\vecr_2 \\\times[1-K(\vecr-\vecr_1+\vecr_2)]U_{t_it_j}(\vecr)p_i(\vecr_1)p_j(\vecr_2);\label{eq:chi_nonid_def}
\end{multline}
see details in Appendix~\ref{sec:deriv1}. Hard-core monomer--monomer repulsion can be absorbed into $U_{tt'}$, so we do not penalize monomer--monomer repulsion explicitly.

To make progress, we assume the following hierarchy of length scales. The pair interaction typically spans a few monomers, so $\ell_U$ is a small multiple of the Kuhn length, $b$. Now, the distance between the $i$th monomer and the chain CM scales as $b$ times $N$ to a positive power~\cite{DoiBOOKpolymer}. For instance, for a Gaussian chain,
\begin{equation}
    p_i(\vecr)=\frac{1}{(2\pi \sigma_i^2)^{3/2}}\exp\left( -\frac{|\vecr|^2}{2\sigma_i^2}\right),\label{eq:SCdist}
\end{equation}
with the mean-squared monomer--CM distance,
\begin{equation}
\frac{3\sigma_i^2}{Nb^2}=\left(\frac iN-\frac{1+1/N}2\right)^2+\frac{1-1/N^2}{12},
\label{eq:MSD}
\end{equation}
so $\sigma_i\sim N^{1/2}b$ for any $i=1,\ldots,N$ (see Fig.~\ref{fig:illust}(c)), representing that monomers near the ends have greater variance in position than monomers near the center of the chain. Since the correlation hole is an ``accumulated'' repulsion from all monomers, we suppose that $K(\vecr)$ is sequence independent and only depends on $N$ through $\ell_\ch$ (\eg $\ell_\ch\sim N^{1/2}b$ for a Gaussian chain). We note that our approach can be readily generalized to non-Gaussian chains~\cite{rubinstein2003polymer}, by replacing the distribution $p(\vecr)$ and correlation-hole kernel $K(\vecr)$ with the appropriate scaling $N^\nu$. Ultimately, we need not know the exact functional form of $K(
\vecr)$, rather its scaling with the chain length.

Therefore, assuming the hierarchy $\ell_U\ll\ell_\ch\ll\sigma_i$, we find an approximate expression for Gaussian chains,
\begin{multline}
    \chi(T)=\frac{z}{2\kt}\frac1{N^2}\sum_{i,j=1}^N\epsilon_{t_it_j}\\\times\left[1-\frac{\int\d\mathbf{x}K(\ell_\ch\mathbf{x})}{(2\pi)^{3/2}}\left(\frac{\ell_\ch^2}{\sigma_{i}^{2}+\sigma_{j}^{2}}\right)^{3/2}\right],\label{eq:chi_nonid_res}
\end{multline}
where $\mathbf{x}$ and $\int\d\mathbf{x}K(\ell_\ch\mathbf{x})$ are dimensionless; see details in Appendix~\ref{sec:deriv2}. The second term in brackets in Eq.~\eqref{eq:chi_nonid_res} embodies the correlation-hole penalty on the pair interaction $\epsilon_{t_it_j}$. Indeed, the further the monomers are from the polymer CM, the larger is $\sigma_i$, the smaller is the penalty, and the larger is the contribution of $\epsilon_{t_it_j}$ to the pair interaction. The position-dependent weights in the summand of Eq.~\eqref{eq:chi_nonid_res} thus impart sequence dependence to $\chi(T)$. 

Equation~\eqref{eq:chi_nonid_res} is our main result\,---\,a minimal, physically interpretable ``intervention'' into the mean-field theory, resulting in a sequence-dependent pair-interaction strength. Our result can be summarized via what we refer to as a residue-accessibility parameter (RAP),
\begin{equation}
    P\equiv \frac1{N^2\bar\epsilon}\sum_{i,j=1}^N\frac{\epsilon_{t_it_j}}{[(\sigma_{i}^{2}+\sigma_{j}^{2})/Nb^2]^{3/2}}\label{eq:RAP},
\end{equation}
which, together with the $N$- and sequence-independent fitting parameter $C_K=[(2\pi N)^{1/2}b]^{-3}\int\d\vecr K(\vecr)$, determine 
\begin{equation}
    \chi(T)=\frac{z\bar\epsilon}{2\kt}(1-C_KP).    
\end{equation}
We later demonstrate that $C_K$ can be parameterized considering only the behavior of homopolymers, which experience the same residue accessibility effect as sequence-resolved polymers. Otherwise, upon setting the temperature scale, there are no other fitting parameters in our approach. We believe that this approach can be readily generalized to non-Gaussian chains, mixtures of different heteropolymers, and microphase separation.

\section{Results}

To assess the importance of residue accessibility, we performed extensive grand-canonical lattice Monte-Carlo simulations of two-letter polymers on a cubic lattice of $z=26$~\cite{Akkiraju2026} (see details in Appendix~\ref{sec:MC}), from which we extracted the critical parameters for phase separation. Each simulation contained a single polymer type defined by its sequence of $\A$- and $\B$-type monomers. These monomers interact via nearest-neighbor interactions
\begin{equation}
    \epsilon_{\A\A}=\frac{1+c}2,\qquad\epsilon_{\A\B}=\epsilon_{\B\A}=\epsilon_{\B\B}=\frac{1-c}2,
\end{equation}
which depend on ``solvent quality'' $c$, characterizing the difference of hydrophobicities for the two residue types~\cite{Panagiotopoulos2024}. The solvent is otherwise implicit. We considered solvent-quality parameters $c=0,1/4,\,1/2,\,3/4,\,1$ and polymer lengths $N=8,\,12,\,14,\,18,\,24$. All sequences with $c=0$ for a given $N$ are degenerate, and so we have counted them as one. Furthermore, since all $\A$- and $\B$-type homopolymers (with $c>0$) for a fixed $N$ only have a single energy scale\,---\,$\epsilon_{\A\A}$ or $\epsilon_\mathrm{\B\B}$, respectively\,---\,by a dimensional argument, the rescaled temperature $\tilde T_\c$ is guaranteed to be the same, so we avoid double-counting those, too. For $N=8$, we simulated all possible $\A$--$\B$ sequence combinations, while for other $N$s we selected many hundreds of sequences (pseudo-) randomly, for a total of 2408 data points. Sampling is designed to span a range of blockiness, as measured by the sequence charge-decoration parameter. From those simulations, we extract the critical temperature that we rescale following Eq.~\eqref{eq:Tctilde}. We note that as $N$ increases, many sequences result in  micelle formation or exhibit microphase separation~\cite{statt2020model, RanaJCP2021}, cases that we do not analyze in this work.

\begin{figure}
\centering
\includegraphics[width=.99\linewidth]{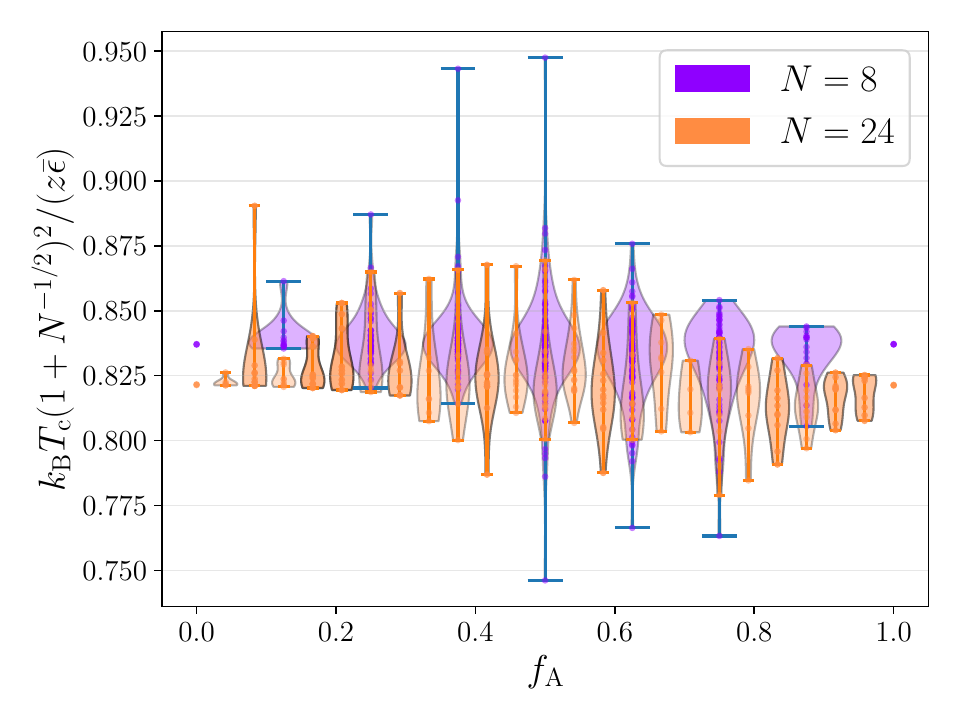}
\caption{Simulated critical temperature $T_\c$ (rescaled according to Eq.~\eqref{eq:Tctilde}) versus the fraction of $\A$-type monomers along a chain, $f_\A$. Were the mean-field prediction, Eq.~\eqref{eq:chi_id}, accurate, we would have found $\tilde{T}_\c=1$ for the entire data set. Clearly, $f_\A$ is insufficient for delineating the sequence dependence of critical temperature.}
\label{fig:data}
\end{figure}

We begin by demonstrating the inadequacy of the conventional FH theory for capturing the sequence dependence of the critical temperature. In Fig.~\ref{fig:data}, we plot the rescaled $\tilde T_\c$ versus the fraction of $\A$-type monomers along a chain, $f_\A=N^{-1}\sum_{i=1}^N\delta_{t_i\A}$. For readability of the figure, we have drawn the data points alongside their spread for only the $N=8,24$ datasets (a total of $639$ points). Were the mean-field FH theory adequate, all points would have been equal to $\tilde T_\c=1$, which is clearly not the case. Notably, the rescaled critical temperature for all chains cluster around $\tilde T_\c\simeq0.83$, implying that the mean-field theory is systematically offset by about $20\%$ for all sequence types.  Furthermore, there is a considerable spread around this constant value with $0.746\leq\tilde{T}_\c\leq 0.947$ for $N=8$.  While the variations in the rescaled $\tilde T_\c$ may appear small, they lead to measurable changes in the dimensional temperature $T_\c$. For example, a $5\%$ change in the predicted value of $\tilde T_\c$ (representing temperatures on the order of magnitude $T_\c\sim300~ \mathrm{K}$ in typical biopolymers~\cite{QuirozNatMat2015,DignonPNAS2018,DignonACS2019}) corresponds to a $15~ \mathrm{K}$ difference in $T_\c$. Therefore, it is crucial to capture sequence dependence to describe the full breadth of critical temperatures.

\begin{figure}
    \centering
    \includegraphics[width=0.99\linewidth]{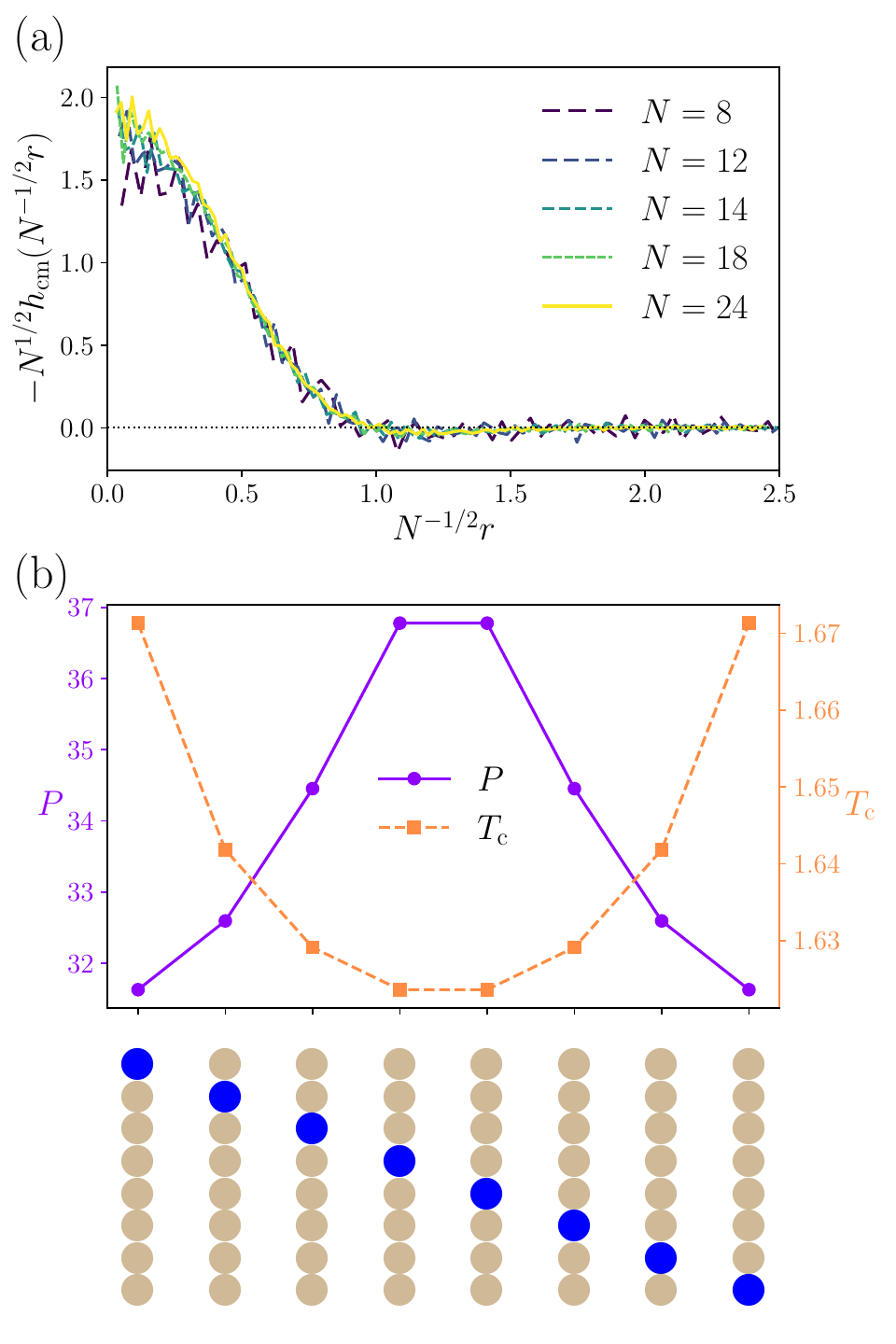}
    \caption{Residue accessibility arises from the correlation hole effect and is correlated with critical temperature. (a) Total correlation function, $h_{\cm}$ among polymer centers of mass for different homopolymer lengths, $N$, computed in the dense phase at $0.7T_\c$, as a function of radial coordinate $r$. The collapse of all curves upon rescaling by $N^{1/2}$ supports our assumption for the universality of a correlation hole. (b) The residue-accessibility parameter (RAP) $P$ and the critical temperature $T_\mathrm{c}$ for $\A\B_7$ chains. ($P$ is given in Eq.~\eqref{eq:RAP}, using $\epsilon_{\A\A}=7/8$ and $\epsilon_{\A\B}=\epsilon_{\B\A}=\epsilon_{\B\B}=1/8$.) The larger RAP is, the less accessible is the attractive residue and the lower is the critical temperature.}
    \label{fig:rdf_and_tp}
\end{figure}

Next, we turn to testing our theoretical proposal. In Fig.~\ref{fig:rdf_and_tp}, we investigate the microscopic basis of these variations of the critical temperature. Fig.~\ref{fig:rdf_and_tp}(a) shows the (negative of the) total correlation function of polymer CMs computed at $T= 0.7 T_\c$ for the homopolymers in the dense phase. We see a marked correlation hole volume whose range scales as $\ell_\ch\sim N^{1/2}$ in units of lattice spacing. This confirms the source for polymer--polymer correlations we used in Sec.~\ref{sec:theory}~\cite{BOOK:deGennes,WangMACRO2017}: If two polymers cannot come into contact with each other, then an attractive residue in the center and one near the end will have very different interaction magnitudes, and therefore should give rise to different critical temperatures. To illustrate this point, in Fig. ~\ref{fig:rdf_and_tp}(b), we plot RAP alongside the (unscaled) $T_\c$ versus the position of the one $\A$-type monomer in an eight-letter chain, with overall composition $\A\B_7$ (\cf Ref.~\cite{panagiotopoulos2024sequence}). Indeed, RAP is negatively correlated with the critical temperature. When RAP is larger for chains with the stronger interacting $\A$-type monomer at the ends, the measured $T_\c$ is lower. This trend tracks with the intuition that stronger effective interactions occur when the residues are more accessible.

\begin{figure}%[tbhp]
\centering
\includegraphics[width=.99\linewidth]{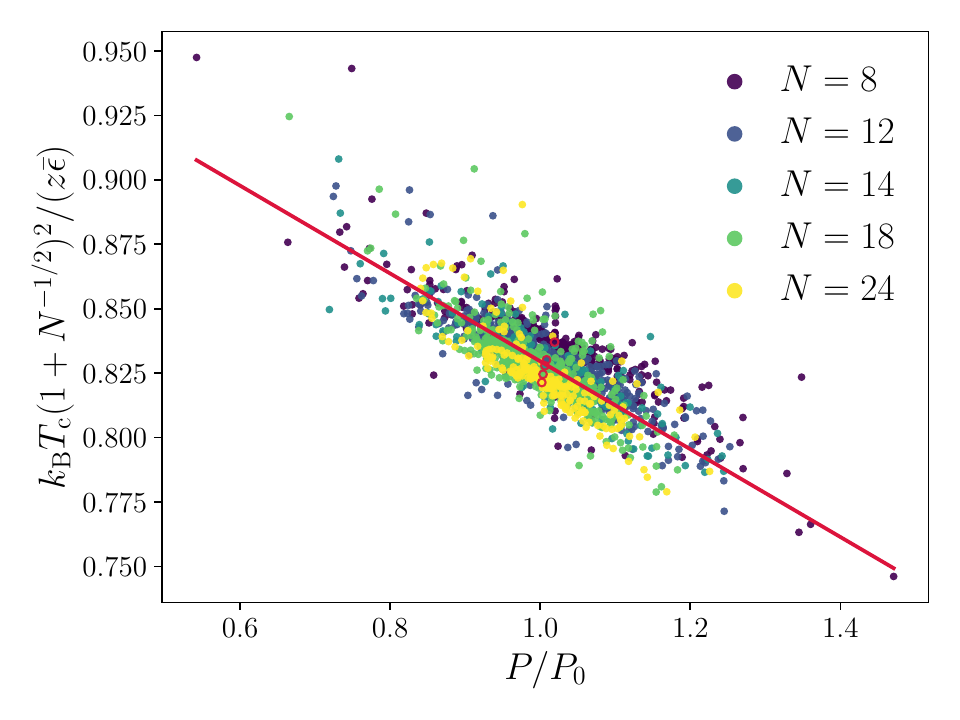}
\caption{Data collapse of Monte-Carlo simulation results for 2408 unique polymer sequences of various lengths (as indicated) $N$ and solvent qualities $c$. Critical temperature $T_\c$ (rescaled according to Eq.~\eqref{eq:Tctilde}) versus RAP (normalized by the constant $P_0\simeq32.76$\,---\,the RAP of a homopolymer, $\epsilon_{ij}=1$, of length $N\to\infty$). The intercept for $P=0$ is fixed at $\tilde{T}_\c=1$. The slope, corresponding to the ratio between the correlation hole and Gaussian-chain radius of gyration, is the only fitting parameter, and is determined by the average of the homopolymer critical temperatures (outlined in red).}
\label{fig:result}
\end{figure}

With these confirmations in hand, we proceed to assess the importance of residue accessibility for the critical temperature. To this end, in Fig.~\ref{fig:result}, we plot our entire simulation dataset\,---\,$2408$ chains of varying lengths, compositions, sequences, and solvent quality\,---\,versus RAP (arbitrarily rescaled by the RAP value of a $c=0$ homopolymer of infinite length, $P_0\simeq32.76$). Indeed, the critical temperatures appear to strongly correlate with RAP for different values of $N$. 

To test our theoretical prediction, Eq.~\eqref{eq:chi_nonid_res}, we draw a line connecting two physically meaningful points: (i) The expected intercept\,---\,if the residues are fully accessible, $P=0$, the mean-field theory (Eq.~\eqref{eq:chi_id}) should be exact, $\tilde T_\c(P=0)=1$. (ii) The rescaled $\tilde T_\c$ for homopolymers found in simulations, which we highlighted in red shown in Fig.~\ref{fig:result}. The principle of decreased accessibility due to the correlation hole should also apply to homopolymers for which indeed $\tilde T_\c\neq1$. Since there is a slight residual $N$-dependence in $\tilde T_\c$, we used the average over the simulated homopolymer datapoints. The line drawn between perfect accessibility at $\tilde{T}_\c(P=0)=1$ and the homopolymer RAP $\tilde{T}_\c(P\simeq P_0)\simeq0.83$ is a reasonable fit to the cloud of rescaled critical temperatures ($R^2=0.63$ for all data points) even though the fit only involves the homopolymers.

There is still residual spread around the fitted line. In fact, the mathematical structure of the FH free energy itself is not perfectly describing a real critical polymer solution, and so we observe a  small downward bias of the ``points cloud'' with increasing $N$. Nonetheless, the spread of points around the fitted prediction also decreases with $N$, which is much smaller compared to that of Fig.~\ref{fig:data}. Overall, we consider the agreement shown in Fig.~\ref{fig:result} encouraging, given that the fitted line only contains information about homopolymers. Therefore, RAP is able to reduce the dimensionality of the problem of sequence dependence from $2^N$ to just $1$ within a small error. We thus envision a computationalist or an experimentalist estimating $T_\c$ for, for example, a homopolymer (or a few heteropolymers), from which they can extrapolate to $T_\c$s of other chain sequences via RAP.

\section{Discussion}

In this work, we studied the sequence dependence of polymer-solution phase diagrams. We propose that the correlation hole~\cite{BOOK:deGennes,WangMACRO2017} between polymer centers of mass causes a nonuniform monomer accessibility. As a result, monomer--monomer interaction strengths depend on the positions of the interacting monomers along their respective chains. This effect is encoded within a residue-accessibility parameter (RAP), which may serve as a first step at delineating the sequence dependence of phase diagrams in sequence-resolved polymer solutions. Indeed, we showed that RAP enables the collapse of critical-temperature data from extensive Monte Carlo simulations of different sequences, compositions, and lengths. Therefore, our approach may allow theories for phase separation as simple as the mean-field FH model to acquire physically interpretable sequence dependence.
 
Prior approaches based on single-chain features inherently do not account for the presence of other polymers in solution. Statistical-field-theory-based computational methods may similarly not capture the influence of the correlation hole or residue accessibility: RPA captures intra-chain correlations, but has not been extended to inter-chain correlations such as the correlation hole~\cite{lin2017charge}, to the best of our knowledge. SCFT can more precisely embed sequence-defined chain conformations, but it is still solved in the mean-field where inter-chain correlations are neglected. 

We thus propose RAP as a collective variable to tackle the curse of dimensionality in the sequence dependence of polymer properties. Given its physical origin, we expect RAP to be most useful for interaction-energy-dependent properties of disordered polymers. We find that RAP does not scale meaningfully with the critical volume fraction (not shown), and it is not guaranteed to capture the sequence dependence of transport coefficients. Likewise, our approach might be too simplistic for more complex systems such as dense polymer suspensions and proteins with well-defined tertiary structures. Moreover, as a single parameter, RAP possesses ``blind spots'' to various sequence-related symmetries by construction.One curious example is that RAP attains the same value for all two-letter chains of the same length that are reverse-complement symmetric (\ie sequences that recover themselves upon reversal and letter exchange, such as $\mathrm{AABBABAABB}$). 

Nonetheless, the precise physical foundation of RAP, stemming from correlation-hole repulsion as a perturbation to a mean-field theory, points toward the extensions required to capture system-specific properties. For instance, inputting the precise single-chain configurational statistics $p_i(\vecr)$ of structured polymers may provide more accurate weights within RAP. Similarly, knowing the precise statistics of a non-Gaussian chain, one may amend the scalings of $p_i(\vecr)$ and $K(\vecr)$ with $N$ accordingly. Furthermore, RAP can readily be extended to mixtures of a few polymer types. Elsewhere, incorporating correlation-hole repulsion into beyond-mean-field approaches might capture the sequence dependence of the critical volume fraction, which has already been related to single-chain statistics such as the sequence charge-decoration parameter~\cite{RanaJCP2021}. Such conceptual and computational extensions would contribute to the challenging task of deciphering the sequence dependence of polymer properties.

% We note that our derivation is phenomenological in character; we assumed the existence of a repulsion kernel $K(\vecr)$ that reduces the probability of overlap of two polymer CMs, and postulated that $K(\vecr)$ scales with distance as the correlation-hole distance $\ell_\ch$. Were we computing the interaction magnitude in the dense phase, the rigorous calculation would simply have amounted to replacing $K(\vecr)$ in Eq.~\eqref{eq:chi_nonid_def} with the total correlation function~\cite{BOOK:Hansen} (that which we have drawn in Fig.~\ref{fig:rdf_and_tp}(b)). As we see, in this case, $h(\vecr)$ should have also scaled with $1/N^{1/2}$ in magnitude~\cite{BOOK:deGennes,WangMACRO2017}, whereas we have supposed in our phenomenological derivation that $K(\vecr)$ only depends on $N$ through $\ell_\ch$.  We are, however, after the critical point, and due to diverging interaction magnitudes, it is unclear how to correctly account for the pair correlations in that regime. In fact, the FH is not constructed to be accurate at the critical point (\eg the entropy is certainly not that of an ideal solution); instead, it is only supposed to extrapolate reasonably well to the critical point~\cite{BOOK:deGennes}.  Future work should look into the correct scaling or perhaps obtain the critical-point pair-interaction magnitudes with the appropriate $h(\vecr)$.

As the polymer lengths and their blockiness in our simulations increased, some polymers have micellized or microphase-separated~\cite{statt2020model, RanaJCP2021}, which we have excluded from our data. This fact has impacted our ability to systematically study sequences whose lengths are much longer. We expect that chain correlations and sequence coherence may drive rich structural features such as microphase separation, stabilization, and micellization that may be represented within our framework with some modifications. Future work will aim to characterize how particular sequence features lead to microphase separation under this theoretical framework. 

Finally, we only explored equilibrium conditions. Interesting sequence effects result in changes in polymer dynamics~\cite{muthukumar2026sticky} and oligemerization kinetics~\cite{haugerud2026theory} in the condensed phase, which may also be influenced by sequence accessibility. We further focused on bulk phase diagrams rather than interfaces or microphase separation. Other interfacial properties such as surface tension have been shown to be sequence dependent  in a way that depends on the blockiness of sequences~\cite{tan2025biomolecular}.  Furthermore, more information-rich, yet expensive, metrics have been identified by using machine-learning approaches, such as the ``expenditure density'' defined by the work of compression of a polymer solution~\cite{oliver2025b} that may intrinsically depend on the accessibility of residues. For more realistic representations of biopolymers, such as intrinsically disordered proteins, the interplay of unstructured and structured regions would lead to stronger correlations that would need to be considered in evaluating sequence-dependent effects. In any of these cases, residue accessibility remains an important consideration for the sequence-dependence of polymer interactions.

\acknowledgements
J.P.D. acknowledges support from the Omenn-Darling Princeton Bioengineering Institute–Innovators (PBI2) Postdoctoral Fellowship. B.S. was supported by the Princeton Center for Theoretical Science and in part by the Center for the Physics of Biological Function at Princeton University. A.A. would like to acknowledge a National Science Foundation Graduate Research Fellowship under Grant No. DGE-2039656. H.A.S. acknowledges support from NSF under Grant No. DMS/NIGMS-2245850. We acknowledge financial support from the Princeton Center for Complex Materials (PCCM), an NSF-supported Materials Research Science and Engineering Center under Award No. DMR-2011750.

J.P.D., B.S., and H.A.S. conceptualized the theoretical model. A.A. and A.Z.P. performed Monte Carlo simulations and developed the model. All authors participated in the writing and drafting of figures.

\appendix

\section{Grand-Canonical lattice Monte Carlo simulation} \label{sec:MC}

More comprehensive descriptions of the lattice chain model~\cite{Larson1985,Larson1988} and the grand-canonical Monte Carlo simulations can be found in previous publications~\cite{PanagiotopoulosMACRO1998, Panagiotopoulos2023, panagiotopoulos2024sequence}. In the following, we provide specific details relevant to the simulations performed for this study. 

Critical points were determined using Monte Carlo simulations in the grand canonical ensemble, where the chemical potential $\mu$, volume $V$, and temperature $T$ are fixed. For a system of $N_\p$ polymer chains of length $N$ in a cubic box of side length $L$, the volume fraction is defined as $\phi = N_\p N/L^3$. Chain insertion and deletion moves were implemented with an athermal Rosenbluth scheme~\cite{Rosenbluth1955} to enhance sampling efficiency.

Critical parameters were extracted using histogram reweighting~\cite{PanagiotopoulosMACRO1998,Panagiotopoulos2000} together with the Ferrenberg–Swendsen method~\cite{Ferrenberg1988}, which enables simulation runs with sufficient overlap in the particle number $N_\p$ and energy $E$ distributions to be merged, thereby reducing the number of separate simulations required. Mixed-field finite-size scaling~\cite{Bruce1992,Wilding1992} was employed, in which the transformed probability distribution of the field-mixed order parameter, $m = n - sE$, is matched to the universal three-dimensional Ising distribution~\cite{Tsypin2000}.

To ensure consistent statistical quality across chain lengths, box sizes were chosen such that $\sim130$ chains were present at the critical volume fraction. This approach provided adequate configurational sampling while limiting finite-size effects and computational cost.

Computer codes for simulations of critical parameters of varying sequences are freely available for download from the Princeton Data Commons repository at \href{https://doi.org/10.34770/gp3g-1m25}{https://doi.org/10.34770/gp3g-1m25}. All data and figure files for the current work are freely available on Github at \href{https://github.com/jpldesouza/residue_accessiblity}{https://github.com/jpldesouza/residue\_accessiblity}.

\section{Derivation: From microscopic pair-interactions to the Flory-Huggins parameter} \label{sec:deriv1}

Formally, the Flory-Huggins free energy can be derived rigorously from statistical-field theory~\cite{BOOK:Frederickson}. In the following, we outline one simple and direct method to obtain Eqs.~\eqref{eq:epsilon} and~\eqref{eq:chi_nonid_def} starting from the pair-interaction energy,
\begin{equation}
    \bar U\equiv\frac12\sum_{n\neq m}^{N_\p}\sum_{i,j=1}^N\langle U_{t_it_j}(\vecr_{n,i}-\vecr_{m,j})\rangle,
\end{equation}
as we have done in the main text. Define the density field of all $i$th monomers, $\rho_i(\vecr)=\sum_{n=1}^N\delta(\vecr-\vecr_{n,i})$, with which the pair interaction energy can be rewritten as
\begin{equation}
    \bar U=\frac12\sum_{i,j=1}^N\int\d\vecr'\int\d\vecr \langle\rho_i(\vecr')\rho_j(\vecr'+\vecr)\rangle U_{t_it_j}(\vecr).\label{eq:Udensity}
\end{equation}
We ignored the correction from self-interaction terms ($n=m$), as they are linear in $\rho_i(\vecr)$ and hence only offset the chemical potential by a constant.

To derive Eq.~\eqref{eq:epsilon}, we consider the ideal limit, where polymers are spatially uncorrelated, $\langle\rho_i(\vecr')\rho_j(\vecr'+\vecr)\rangle\simeq\langle\rho_i(\vecr')\rangle\langle\rho_j(\vecr'+\vecr)\rangle$, so Eq.~\eqref{eq:Udensity} becomes
\begin{equation}
    \bar U=\frac12\sum_{i,j=1}^N\int\d\vecr'\int\d\vecr \langle\rho_i(\vecr')\rangle\langle \rho_j(\vecr'+\vecr)\rangle U_{t_it_j}(\vecr).
\end{equation}
During bulk phase separation, the $\chi$ parameter is computed from bulk pair-interactions, so we consider a uniform density in space, $\langle\rho_i(\vecr)\rangle=N_\p/V$. This yields
\begin{equation}
    \bar U=\frac{N_\p^2}{2V}\sum_{i,j=1}^N\int\d\vecr U_{t_it_j}(\vecr),
\end{equation}
where we used $\int\d\vecr'=V$ as no other quantity in the integrand depended on $\vecr'$. Recall the definition of the total polymer volume fraction, $\phi=N_\p N\upsilon_\m/V$, with which we obtain
\begin{equation}
    \frac{\upsilon_\m\bar U}{V\kt}=-\frac1{2\kt}\frac1{N^2}\sum_{i,j=1}^N\left[-\frac1{\upsilon_\m}\int\d\vecr U_{t_it_j}(\vecr)\right].
\end{equation}
This indeed coincides with $\upsilon_\m\bar U/(V\kt)=-\chi\phi^2$, where $\chi$ is given in Eq.~\eqref{eq:chi_id} and $\epsilon_{tt'}$ in Eq.~\eqref{eq:epsilon}.

To derive Eq.~\eqref{eq:chi_nonid_def}, we must not assume independent polymers, as they are correlated via the correlation hole. In the main text, we approximated the polymer internal structures to be independent, as if the correlation hole occurs via the polymer CMs. Thus, we first decompose the $i$th monomer density field in terms of the polymer CM density field, $\rho^\cm(\vecr)$, convolved with the (independent) single-chain monomer--polymer CM distance distribution, $p_i(\vecr)$ (Eq.~\eqref{eq:SCdist}),
\begin{equation}
    \rho_i(\vecr)=\int \d\vecr_1\rho^\cm(\vecr-\vecr_1)p_i(\vecr_1).
\end{equation}
Inserting in Eq.~\eqref{eq:Udensity}, we get
\begin{multline}
    \bar U=\frac12\sum_{i,j=1}^N\int\d\vecr'\int\d\vecr \int\d\vecr_1\int\d\vecr_2\\\times\langle\rho^\cm(\vecr'+\vecr_1)\rho^\cm(\vecr'+\vecr+\vecr_2)\rangle U_{t_it_j}(\vecr)p_i(\vecr_1)p_j(\vecr_2),\label{eq:UdensityCM}
\end{multline}
where, as pointed out above, we assume that $p_i(\vecr)$ are independent of other polymers, so the correlations are only communicated via the polymer CM positions. Normalizing by the polymer density, $N_\p/V$, we defined our phenomenological repulsion kernel as 
\begin{equation}
    \langle\rho^\cm(\vecr)\rho^\cm(\vecr')\rangle\equiv\frac{N_\p^2}{V^2}[1-K(\vecr-\vecr')].
\end{equation}
Inserting this in Eq.~\eqref{eq:UdensityCM}, once again no quantity depends on $\vecr'$, so we find
\begin{multline}
    \bar U=\frac{N_\p^2}{2V}\sum_{i,j=1}^N\int\d\vecr \int\d\vecr_1\int\d\vecr_2\\\times [1-K(\vecr-\vecr_1+\vecr_2)]U_{t_it_j}(\vecr)p_i(\vecr_1)p_j(\vecr_2).\label{eq:UdensityCM2}
\end{multline}
With that, we obtained Eq.~\eqref{eq:chi_nonid_def} with $-\chi\phi^2=\upsilon_\m\bar U/(V\kt)$.\linebreak

\section{Derivation: Modified Flory-Huggins parameter}\label{sec:deriv2}

Equipped with Eq.~\eqref{eq:chi_nonid_def}, we show in the following how we obtained Eq.~\eqref{eq:chi_nonid_res} using the assumed hierarchy $\ell_U\ll\ell_\ch\ll\sigma_i$ and Eq.~\eqref{eq:SCdist}. This separation of $\ell_\mathrm{ch}$ and $\sigma_i$ is intended as a leading order correction, since due to their similar scaling with chain length, the separation of length scales may not be explicitly valid, especially for central monomers. 

We first use the normalization of $p_i(\vecr)$ and redefine $\vecr_2=\vecr_1-\vecr+\vecr'$, with which Eq.~\eqref{eq:chi_nonid_def} becomes
\begin{multline}
    \chi(T)=-\frac1{2v_\m\kt}\frac1{N^2}\sum_{i,j=1}^N\left[\int\d\vecr U_{t_it_j}(\vecr)\right.\\\left.-\int\d\vecr'K(\vecr')\int\d\vecr U_{t_it_j}(\vecr)\int\d\vecr_1p_i(\vecr_1)p_j(\vecr_1-\vecr+\vecr')\right].\\
\end{multline}
Now, since $K(\vecr)$, $U(\vecr)$, and $p_i(\vecr)$ tend to $0$ after a distance of order $\ell_\ch$, $\ell_U$, and $\sigma_i$, respectively, we note that only distances  $|\mathbf{r}'|\sim\ell_\ch$, $|\vecr|\sim\ell_U$, and $|\vecr_1|\sim\sigma_i$ contribute to the integral. Owing to the hierarchy $\ell_U\ll\ell_\ch\ll\sigma_i$, we neglect $\vecr$ and $\vecr'$ compared to $\vecr_1$, so
\begin{multline}
    \chi(T)=\frac z{2\kt}\frac1{N^2}\sum_{i,j=1}^N\epsilon_{t_it_j}\\\times\left\{1-\left[\int\d\vecr'K(\vecr')\right]\int\d\vecr_1p_i(\vecr_1)p_j(\vecr_1)\right\}.\label{eq:chi_deriv}
\end{multline}
where we used Eq.~\eqref{eq:epsilon} to substitute $\int\d\vecr U_{tt'}(\vecr)$. (Since both $\sigma_i$ and $\ell_\ch$ scale as $N^{1/2}b$, it may be that the proportionality coefficient in $\ell_\ch$ is not much smaller than that of $\sigma_i$. In this case, one should regard Eq.~\eqref{eq:chi_deriv}  (and Eq.~\eqref{eq:chi_nonid_res}) as the leading-order result.) Using the Guassian-chain statistics, Eq.~\eqref{eq:SCdist}, 
\begin{equation}
    \int\d\vecr_1p_i(\vecr_1)p_j(\vecr_1)=[2\pi(\sigma_i^2+\sigma_j^2)]^{-3/2},
\end{equation}
with which we recover Eq.~\eqref{eq:chi_nonid_res}.

\end{document}